\documentclass[showpacs,twocolumn,prl,aps]{revtex4}
\usepackage{graphicx}

\begin{document}

\title[Short title for running header]{Antiferromagnetic long range order in the uniform resonating valence bond state on square lattice}
\author{Tao Li}
\affiliation{ Department of Physics, Renmin University of China,
Beijing 100872, P.R.China}
\date{\today}

\begin{abstract}
With extensive variational Monte Carlo simulation, we show that the
uniform resonating valence bond state(U-RVB) on square lattice is
actually antiferromagnetic long range ordered. The ordered moment is
estimated to be $m\approx 0.17$. Finite size scaling analysis on
lattice up to lattice size of $50\times 50$ shows that the spin
structure factor at the antiferromagnetic ordering wave vector
follows perfectly the $\mathrm{S}_{\mathrm{q}=(\pi,\pi)} \simeq
S_{0}+\alpha (1/L)^{\frac{5}{4}}$ behavior, where $L$ is linear
scale of the lattice. Such a behavior is quite unexpected from the
slave Boson mean field treatment or the Gutzwiller approximation of
the uniform RVB state.
\end{abstract}
\maketitle

The resonating valence bond(RVB) state plays an important role in
our understanding of exotic physics in quantum antiferromagnet and
the high T$_{c}$ superconductors\cite{RVB}. The RVB state is a
quantum many body spin singlet constructed from coherent
superposition of different pairing patterns of the spins on the
lattice.

In practice, the most studied RVB states are those generated from
Gutzwiller projection of Fermionic or Bosonic mean field state with
condensed singlet pairs. It is well known that the RVB states
generated from these two routes differ in their property
qualitatively. For example, a Bosonic RVB state can be either short
ranged in its spin correlation when the spinon excitation spectrum
has a full gap, or , long range ordered when the Bosonic spinon
condense. On the other hand, a Fermionic RVB state can have more
choices for its spin correlation. Beside being short ranged with an
exponentially decaying spin correlation, the spin can also choose to
stay in various kind of critical states. For example, a Fermionic
RVB state can have a large spinon Fermi surface or various kind of
node structure in its spinon dispersion. The spin correlation in
such critical state generally follows power law behavior. However,
it is generally believed that it is difficult to describe state with
long range magnetic order within the Fermionic RVB scheme without
breaking the spin rotational symmetry. Such a difference between the
Bosonic RVB state and the Fermionic RVB state stems from the fact
that the magnetic long range order in the Bosonic RVB state actually
originates from the condensation of the Bosonic spinon, while in the
Fermionic RVB state, the Fermionic spinon must be paired before
condensation. However, a condensation of triplet pair will
inevitably break the spin rotational symmetry.

The relation between the Bosonic and Fermionic RVB state has been
studied by many authors\cite{Sorella,Read}. It is found that when
the RVB amplitude is extremely short ranged, or, extending only
between nearest neighboring sites, and the lattice is planar, a
Fermioinc RVB state can be transformed into a Bosonic RVB state with
suitable redefinition of the RVB amplitudes.  For Fermionic RVB
state with longer range RVB amplitudes, their relation with Bosonic
RVB state is still elusive. More generally, it is found that on the
bipartite lattice, both Bosonic and Fermionic RVB state satisfy the
well known Marshall sign rule for unfrustrated antiferromagnet, if
they are both derived from a bipartite mean field Hamiltonian.

An analytical treatment of the RVB state is difficult. The most
commonly adopted method to study the Fermionic RVB state is the
slave Boson mean field theory. Beyond the mean field theory, there
is also the more elaborate effective gauge field theory
treatment\cite{gauge,Mudry,Kim}. However, it is well known that the
mean field theory may fail qualitatively as the no double occupancy
constraint is relaxed to a global one in such treatment. The gauge
fluctuation correction, which can in principle improve the result of
the mean field theory, are accounted for only at the Gaussian level
in most cases and missed the singular gauge fluctuation effect at
the lattice scale. As a variational state, the Fermionic RVB state
is also studied by the Gutzwiller approximation\cite{GA}, which try
to estimate the effect of the local constraint by correcting the
mean field expectation value with some correction factor. However,
the Gutzwiller approximation can not produce any qualitatively
different prediction from the mean field theory as to the spin
correlation in the RVB state\cite{Hetanyi}.

The variational Monte Carlo(VMC) method is a direct way to study the
properties of the Fermionic RVB state. It is found that in certain
cases the VMC result can be quite different from the mean field
prediction. For example, on bipartite lattice, a well defined sign
structure(the Marshall sign rule) may emerge out of a general mean
field state through the Gutzwiller projection, even if the mean
field state breaks the time reversal symmetry and has a complex
valued wave function. Such a sign structure will restore the time
reversal symmetry in the projected state and remove from the system
possible topological degeneracy.

In this paper, we present another example in which the mean field
prediction is qualitatively changed by the Gutzwiller projection. We
show the well know uniform RVB state on the square lattice with a
large Fermi surface for the spinon actually describes a state with
magnetic long range order.

The uniform RVB state first appears as a variational state for the
high-Tc superconductors in Anderson's original paper on RVB theory
of the cuprates\cite{BZA}. The state has a large nested Fermi
surface for the Fermionic spinon. Although it is later found that
the d-wave RVB state is energetically more favorable than the
uniform RVB state\cite{d-wave}, the uniform RVB state is
nevertheless still an interesting state of its own right.

At the mean field level, the uniform RVB state is described by a
filled Fermi sea with a nested Fermi surface. The spin correlation
function decays algebraically with distant as $\frac{1}{R^{4}}$. The
spin structure factor at the antiferromagnetic wave vector is given
by $S_{q=(\pi,\pi)}=\frac{3}{2N}$, in which $N$ is number of lattice
sites\cite{Arovas}. Thus at the mean field level, the uniform RVB
state describes a state with no magnetic long range order. However,
as the system presents a nested Fermi surface, the mean field state
is instable with respect to infinite small perturbation toward
antiferromagnetic ordering. Thus it is highly possible that the
Gutzwiller projection may induce finite antiferromagnetic order in
the uniform RVB state, in a way that the spin rotational symmetry
remains intact.

The uniform RVB state is given by the Gutzwiller projection of the
ground state of the following mean field ansatz
\begin{equation}
\mathrm{H}_{U-\mathrm{RVB}}=-\sum_{\langle i,j
\rangle,\sigma}(c^{\dagger}_{i,\sigma}c_{j,\sigma}+h.c.),
\end{equation}
whose mean field dispersion is given by
$\epsilon_{\mathrm{k}}=-2(\cos\mathrm{k}_{x}+\cos\mathrm{k}_{y})$.
The uniform RVB state can then be written as
\begin{equation}
|U-\mathrm{RVB}\rangle=\mathrm{P}_{\mathrm{G}}\prod_{\mathrm{k}<\mathrm{k}_{\mathrm{F}}}c^{\dagger}_{\mathrm{k}\uparrow}c^{\dagger}_{\mathrm{-k}\downarrow}|0\rangle,
\end{equation}
where $\mathrm{k}_{\mathrm{F}}$ denotes the Fermi momentum of the
spinon and is determined by $\epsilon_{\mathrm{k}_{\mathrm{F}}}=0$.
$\mathrm{P}_{\mathrm{G}}$ denotes the Gutzwiller projection into the
subspace of no double occupancy. The state can also be written in
the form of condensed pairs,
\begin{equation}
|U-\mathrm{RVB}\rangle=\mathrm{P}_{\mathrm{G}}\left(\sum_{i,j}a(i-j)c^{\dagger}_{i\uparrow}c^{\dagger}_{j\downarrow}\right)^{\frac{N}{2}}|0\rangle,
\end{equation}
in which
$a(i-j)=\sum_{\mathrm{k}<\mathrm{k}_{\mathrm{F}}}e^{i\mathrm{k}\cdot(i-j)}$.
Due to the bipartite nature of the square lattice, it can shown that
the uniform RVB state satisfy the Marshall sign rule for a
unfrustrated antiferromagnet\cite{Sorella,Li}. More specifically,
the wave function of the uniform RVB state in a Ising basis is real
up to a global phase factor and its sign is given by
$(-1)^{N^{A}_{\downarrow}}$, where $N^{A}_{\downarrow}$ denotes the
number of down spins in A sublattice.

To detect possible magnetic long range order in the uniform RVB
state, we focus on the spin structure factor at the
antiferromagnetic ordering wave vector
\begin{equation}
S_{\mathrm{q}=(\pi,\pi)}=\frac{1}{N^{2}}\sum_{i,j}e^{i\mathrm{q}\cdot(i-j)}\langle
S_{i}\cdot S_{j}\rangle.
\end{equation}
At the mean field level, the spin structure factor can be found to
be given by $S_{q=(\pi,\pi)}=\frac{3}{4N}$. On the other hand, when
the no double occupancy constraint is enforced by the Gutzwiller
projection, the result is quite different.

\begin{figure}[h!]
\includegraphics[width=8cm,angle=0]{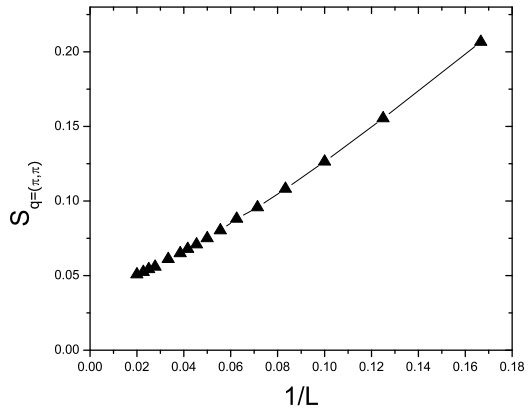}
\caption{The spin structure factor of the uniform RVB state at
$\mathrm{q}=(\pi,\pi)$ on a $L\times L$ square lattice as a function
of the inverse linear scale of the lattice, $1/L$. Presented in this
figure is the results for
$L=6,8,10,12,14,16,18,20,22,24,26,30,36,40,44$ and $50$.
Periodic-antiperiodic boundary condition is adopted in the
calculation to satisfy the closed shell condition. The error bars
are smaller than the size of the symbols.} \label{fig1}
\end{figure}

In Figure 1, we show the VMC result for the spin structure factor
for lattice with size up to $50\times 50$ as a function of the
inverse linear scale of the lattice, $1/L$. In our calculation, we
have used the periodic-antiperiodic boundary condition to satisfy
the closed shell condition. We have used up to $4\times 10^{6}$
statistically independent samples to estimate the spin structure
factor. Each sample is drawn from the Markov chain with $N$ steps of
local updates. The statistical error estimated from the data is
smaller than the size of the symbols in the figure. The spin
structure factor of the uniform RVB state clearly extrapolates to a
finite value in the thermodynamic limit.

To see the trend of the spin structure factor more clearly, we fit
the data with the formula
$S_{q=(\pi,\pi)}=S_{0}+\alpha(1/L)^{\beta}$. The best fit to the
data is found to be $\beta=\frac{5}{4}$ and $S_{0}=0.038$. Thus the
ordered moment in the uniform RVB state is given by
$m=\sqrt{S_{0}}=0.17$, which is quite considerable as compared to
the prediction of the spin wave theory for the Heisenberg model on
square lattice. In Figure 2, a linear fit to the spin structure with
$(1/L)^{\frac{5}{4}}$ as the coordinates is shown. The linear fit is
seen to work extremely well from the smallest lattice we have
studied ($6\times 6$) directly to the largest lattice $50\times50$.

\begin{figure}[h!]
\includegraphics[width=8cm,angle=0]{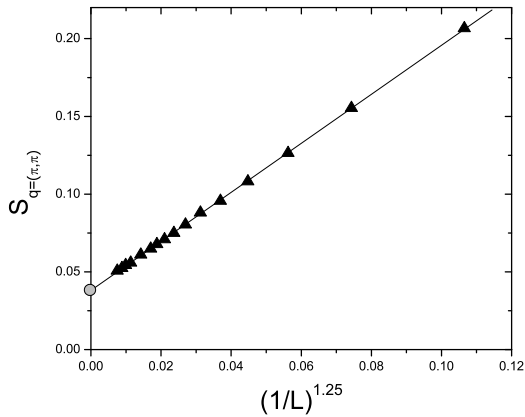}
\caption{The spin structure factor of the uniform RVB state at
$\mathrm{q}=(\pi,\pi)$ on a $L\times L$ square lattice as a function
of $(1/L)^{\frac{5}{4}}$. Presented in this figure is the results
for $L=6,8,10,12,14,16,18,20,22,24,26,30,36,40,44$ and $50$.
Periodic-antiperiodic boundary condition is adopted in the
calculation to satisfy the closed shell condition. The gray dot on
the y-axis denotes the fitted value of $S_{0}$. The error bars are
smaller than the size of the symbols.} \label{fig2}
\end{figure}

As a comparison, we also present the spin structure factor of
another well known Fermionic RVB state on square lattice, namely,
the $\pi$ flux phase in Figure 3. The $\pi$-flux phase is generated
by the following mean field ansatz
\begin{equation}
\mathrm{H}_{\pi-flux}=-\sum_{\langle i,j
\rangle,\sigma}(e^{i\phi_{i,j}}c^{\dagger}_{i,\sigma}c_{j,\sigma}+h.c.),
\end{equation}
in which the phase factor $\phi_{i,j}$ is introduced to guarantee
that each plaquette of the square lattice is threaded by a $U(1)$
flux of value $\pi$. In the $\pi$-flux phase, the spinon has a Dirac
type linear dispersion and the mean field spin correlation also
decay with distance as $1/R^{4}$. After the Gutzwiller projection,
the spin correlation of the system is greatly enhanced. But as is
clear in Fig.3, the $\pi$-flux phase does support an
antiferromagnetic long range order. In fact, a finite size scaling
analysis shows that the spin structure factor decays as
$1/L^{\frac{3}{2}}$ with the linear scale of the lattice and
extrapolates to zero in the thermodynamic limit(Fig. 4).
\begin{figure}[h!]
\includegraphics[width=8cm,angle=0]{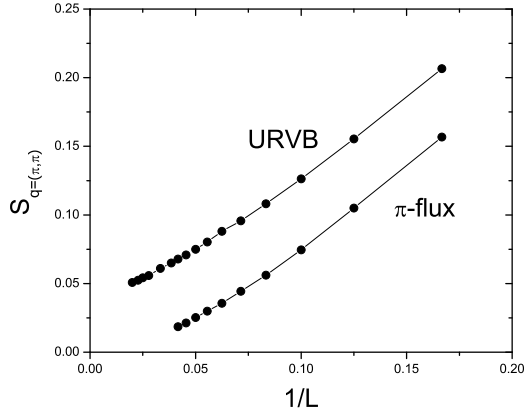}
\caption{The spin structure factor of the $\pi$-flux phase at
$\mathrm{q}=(\pi,\pi)$ on a $L\times L$ square lattice as a function
of the inverse linear scale of the lattice as compared to the result
of the uniform RVB state. Presented in this figure is the results
for $L=6,8,10,12,14,16,18,20,22$ and $24$. Periodic-antiperiodic
boundary condition is adopted in the calculation to satisfy the
closed shell condition. The error bars are smaller than the size of
the symbols. } \label{fig3}
\end{figure}

\begin{figure}[h!]
\includegraphics[width=8cm,angle=0]{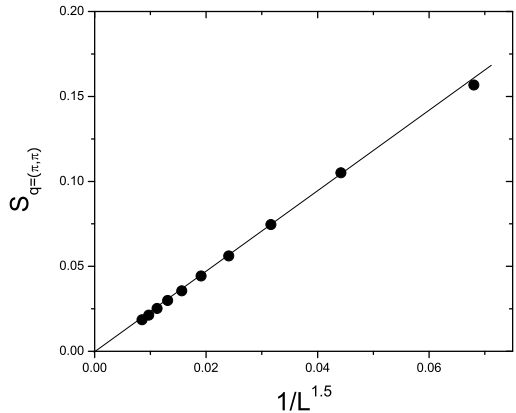}
\caption{The spin structure factor of the $\pi$-flux phase at
$\mathrm{q}=(\pi,\pi)$ on a $L\times L$ square lattice as a function
of $1/L^{1.5}$. Presented in this figure is the results for
$L=6,8,10,12,14,16,18,20,22$ and $24$. Periodic-antiperiodic
boundary condition is adopted in the calculation to satisfy the
closed shell condition. The error bars are smaller than the size of
the symbols.} \label{fig4}
\end{figure}

From these data, it is clear that uniform RVB state indeed describes
a state with antiferromagnetic long range order. This is quite
unexpected from the mean field theory and it represents the first
example of Fermionic RVB state with magnetic order. Such a order
does not originate from the condensation of spinon as in the Bosonic
RVB state and respect the spin rotational symmetry. In the uniform
RVB state, the magnetic long range order can be understood as a
result of the induced magnetic ordering in a system with nested
Fermi surface.

Following the above reasoning, it is quite interesting to ask the
following question: does the uniform RVB state have
antiferromagnetic long range order in other spatial dimension. In
one dimension, the projected Fermi sea state is critical and the
spin correlation function decay with distance as $\frac{1}{R}$,
indicating no magnetic long range order. This can be taken as the
result of the strong quantum fluctuation in one dimensional system.
As fluctuation becomes weaker in higher spatial dimension, it is
quite possible that the uniform RVB state in three dimension(on
cubic lattice) also have antiferromagnetic long range order. A study
of this issue, which is numerically quite demanding for large
lattice, will be presented in a separated paper.

We note a recent preprint\cite{Zhang} independently confirmed the
results of this manuscript.

This work is supported by NSFC Grant No. 10774187 and National Basic
Research Program of China No. 2007CB925001 and No. 2010CB923004.

\end{document}